\documentstyle[12pt]{article}
\begin{document}
\begin{center}
{\bf Progress in the Development of Global Medium-Energy \\
Nucleon-Nucleus Optical Model Potentials} \\
\vspace{12pt}
D.\ G.\ Madland\\
\vspace{12pt}
{\it Theoretical Division, Los Alamos National Laboratory\\
Los Alamos, New Mexico 87545}
\end{center}
 
\begin{abstract}
\noindent Two existing global medium-energy nucleon-nucleus phenomenological
optical
model potentials are described and compared with experiment
and with each other.
The first of these employs a Dirac approach (second-order reduction) that is
global in projectile energy and projectile isospin and applies
to the target nucleus $^{208}$Pb.
Here the standard S-V (isoscalar-scalar, isoscalar-vector) model has been
extended to include the corresponding isovector components by introduction
of a relativistic Lane model.
The determination of the energy range, energy dependence, and isospin
dependence are discussed, as are the predictions for neutron scattering
observables, and also the correlations and ambiguities found in Dirac
phenomenology.
The second of these employs a relativistic equivalent to the Schr\"odinger
equation (including
relativistic kinematics) that is global in projectile energy, projectile
isospin, and target (Z,A).
Here, particular attention is given to predictions for the integrated
scattering
observables -- neutron total cross sections and proton total reaction cross
sections -- and their sensitivity to the absorptive parts of the potential.
Finally, current work is described and
the influence of the nuclear bound state problem (treated in
relativistic mean field theory) on the Dirac scattering problem is
mentioned. Spherical target nuclei are
treated in the present work and strongly-collective target nuclei
(rotational and vibrational) requiring coupled-channels approaches will
be treated in a future paper.
\end{abstract}
 
\begin{center}
{\bf A Global Phenomenological Dirac Potential} \\
\end{center}
\noindent The potential described in this section consists of a
global medium-energy nucleon-nucleus phenomenological Dirac potential for
the target nucleus $^{208}$Pb. The potential is global in projectile energy
and projectile isospin and it was determined \cite{KM89} by least-squares
adjustment of
calculated scattering observables (model parameters) with respect to
corresponding measured scattering observables for both proton and neutron
scattering over a wide range in projectile energy.
 
\vspace{10pt}
\noindent The Dirac equation is used in the mean field approximation by which
the nucleon (meson) fields are replaced by their expectation values.
Proton-nucleus (or neutron-nucleus) scattering is then described using
isoscalar-scalar and isoscalar-vector mean fields. Here these are taken,
respectively, as a spherically symmetric complex Lorentz scalar potential
$S_{0}(r,E, ...)$ corresponding to the (fictitious) $\sigma$ meson field and
a spherically symmetric complex Lorentz vector potential $V_{0}(r,E,...)$
(time-like component
of Lorentz four-vector) corresponding to the $\omega$ meson field, together
with a spherically symmetric Coulomb potential $V_{c}$.
 
\vspace{10pt}
\noindent However, a description of {\it nucleon-nucleus} scattering requires
the explicit
addition of isovector-scalar and isovector-vector potentials (mean fields)
$S_{1}(r,E,...)$
and $V_{1}(r,E,...)$, respectively, yielding
\begin{equation}
  S = S_{0} \pm \epsilon\:S_{1}
\end{equation}
\begin{equation}
  V = V_{0} \pm \epsilon\:V_{1}
\end{equation}
\begin{equation}
  \epsilon = 4\:\vec{T}\cdot\vec{\tau}/A = (N - Z)/A \; .
\end{equation}
In these equations $S_{1}$ and $V_{1}$ correspond to $\delta$ meson and
$\rho$ meson mean fields, respectively, and we use the nuclear physics
isospin convention: $\tau_{3}(neutron) = + \frac{1}{2}\:, \tau_{3}(proton) = -
\frac{1}{2}\:.$ Equations (1)--(3) are a relativistic generalization of the
Lane model \cite{LA62}.
 
\vspace{10pt}
\noindent With this scalar-vector interaction including isospin the Dirac
equation
becomes ($\hbar = c = 1$)
\begin{equation}
  [\vec{\alpha}\cdot\vec{p} + \beta\{m + S\}]\psi = [E - V -V_{c}]\psi
\end{equation}
where $\psi$ is a four-component Dirac spinor with upper and lower components
$\psi_{U}$ and $\psi_{L}\:,\;E$ is the total energy of the scattered nucleon
in the c.m. frame, $\vec{\alpha}$ and $\beta$ are four Hermitian operators
acting on the spin variables alone (these are related to the Dirac $\gamma$
matrices), and $\psi$ contains a two-component isospinor which is an
eigenvector
of $\tau_{3}$ appearing in $S$ and $V$.
A second-order reduction for the upper component $\psi_{U}$ yields
\begin{equation}
 [p^{2} + U_{c} + U_{so}\{(\vec{\sigma}\cdot\vec{L}) -i(\vec{r}\cdot\vec{p}
)\}]
\psi_{U} = [(E - V_{c})^{2} - m^{2}]\psi_{U}
\end{equation}
where the effective central potential $U_{c}$ is given by \\
\begin{equation}
  U_{c} = [2\:E\:V + 2\:m\:S - V^{2} + S^{2} + U_{cc}]/2\:E
\end{equation}
the Coulomb correction term $U_{cc}$ (numerator) is \\
\begin{equation}
  U_{cc} = - 2\:V_{c}\:V
\end{equation}
and the spin-orbit term $U_{so}$ is \\
\begin{equation}
  U_{so} = - \frac{1}{2E}\left\{\frac{1}{r}\:\frac{1}{E + m + S - V -V_{c}}\:
\frac{\partial}{{\partial}r}(S - V - V_{c})\right\} \; .
\end{equation}
 
\vspace{10pt}
\noindent It is worth noting that $S$ and $V$ appear both linearly and
quadratically
in the effective central potential $U_{c}$
leading {\it naturally} to the ``wine-bottle''shapes required to describe
medium-energy nucleon-nucleus scattering somewhat below the transition region
(where
the sign of $U_{c}$ changes) \cite{Ar81}.
Also, the Coulomb correction and spin-orbit terms both appear {\it naturally}
in the Dirac formalism whereas they are {\it ad hoc} in the Schr\"odinger
formalism.
 
\vspace{10pt}
\noindent Equation (5) is solved for the extensive $^{208}$Pb data set by
making
the following assumptions (due to {\it tractability} and the fact that there
exists much more proton data than neutron data): (1) the geometries of the
potentials are independent of projectile species and projectile energy,
so that
all energy dependence and isospin dependence is contained in the strengths of
the potentials, and (2) the same geometry exists for the isoscalar
and isovector components of a given potential.
With these assumptions
\begin{equation}
U = U_{0}(T,\epsilon)\:g(r)
\end{equation}
where $U_{0}$ is a strength, $T$ is the projectile kinetic energy in the
laboratory system, and $g(r)$ is a geometric form factor taken to be a
symmetrized Woods-Saxon shape (which has a closed-form Fourier transform)
given by
\begin{equation}
g(r) = [1 + exp\left(\frac{r - c}{a}\right)]^{-1}\:[1 + exp\left(-\frac{r +
c}{a}
\right)]^{-1}
\end{equation}
where $c$ and $a$ are the radius and diffuseness parameters, respectively, and
$c$ is assumed to be of the usual form $c = r_{0}\:A^{\frac{1}{3}}$ with
$r_{0}$ constant and $A$ the target mass number.
 
\vspace{10pt}
\noindent Energy dependence was studied by considering p + $^{208}$Pb
scattering
data {\it only} which implies that $U_{0}(T,\epsilon) = U_{0}(T)\:,$ and
six forms of $U_{0}(T)$ were tested :
\begin{eqnarray}
U_{0}(T) & = & U_{0} \\
& = & U_{0} + \alpha\:T \\
& = & U_{0} + \alpha\:ln(T) \\
& = & U_{0}\:exp(-T/\alpha) \\
& = & U_{0}\:[1 + (T/\alpha)^{2}]^{-1} \\
& = & U_{0}\:[1 + (T/\alpha)^{2}]^{-\frac{1}{2}}
\end{eqnarray}
\noindent The measured proton scattering observables used in studying the
energy dependence consist of differential elastic scattering cross
sections $d\sigma/d\Omega$, analyzing powers $A_{y}(\theta)$, spin-rotation
functions $Q(\theta)$, and total reaction cross sections $\sigma_{R}$, all as
a function of laboratory proton energy $T_{p}$ for the $^{208}$Pb target.
Considering experimental data over the range 80 to 800 MeV the minimum values
of chi-square/point/energy (data set) are shown in Fig. 1 for four of the six
energy dependencies chosen for study [the other two, Eqs. (15) and (16),
yielded
poorer results, on average, than the linear, log, or exponential energy
dependencies]. The figure shows that no energy dependence is inadmissable
and that none of the three energy dependencies shown is admissable over the
entire energy range shown. The results of reducing the proton energy range
are shown in Fig. 2 for these three energy dependencies. Clearly, a factor
$\sim$ 5 improvement in the total chi-square/point is obtained by reducing the
proton energy range from 80--800 MeV to 95--300 MeV. Note, however, that
the total
chi-square/point is almost equivalent for the energy range of 95--500 MeV.
The remainder of this section addresses the smallest of these three energy
ranges:
95--300 MeV.
 
\vspace{10pt}
\noindent The isospin dependence was studied by including the n + $^{208}$Pb
scattering data (consisting of neutron total cross sections $\sigma_{T}$
as a function of laboratory neutron energy $T_{n}$) with the proton data, for
the energy range 95--300 MeV. Two energy dependencies were chosen to study
the isospin dependence. These are the logarithmic energy dependence,
Eq. (13), which
yields the best chi-square/point of all cases studied and has an
energy-independent
isospin dependence by construction (as does the linear assumption which has a
slightly worse chi-square/point) and the exponential energy dependence, Eq.
(14), which
has an energy dependent isospin dependence by construction.
Thus, the two potential strengths tested are
of the form
\begin{eqnarray}
U_{0}(T,\epsilon) & = & B \pm \epsilon\:C + \alpha\:ln(T) \\
& = & [B \pm \epsilon\:C]\;exp(-T/\alpha)
\end{eqnarray}
where $B, \; C$, and $\alpha$ are the constants to be determined for each of
the four terms of the complete complex scalar-vector interaction potential.
The chi-square minimization led to a logarithmic model that gives slightly
better fits to the neutron data ($\chi^{2}_{tot}(n)$/point of 0.90 {\it vs}
0.95)
and the proton data ($\chi^{2}_{tot}(p)$/point of 12 {\it vs} 13) than the
exponential model and a total-chi square, $\chi^{2}_{tot}$/point, for combined
neutron and proton data, of 12.0 for the logarithmic model {\it vs}
12.8 for the exponential model. The best-fit potentials for these two choices
of the energy and isospin dependence, Eqs. (17) and (18), are given in Table I.
 
\medskip
\begin{center}
\bf{Table I. Best-fit Dirac global optical potentials for nucleon plus
$^{208}$Pb \\ scattering
in the energy interval 95 $\mbox{\boldmath $\leq\:T\:\leq$}$ 300 MeV.\protect
\footnote{Strengths are
in MeV and geometry is in fm; the upper (lower) signs refer to neutrons
(protons).}} \\
\vspace{12pt}
\begin{tabular}{|l|lr|lr|} \hline
&\multicolumn{2}{|c|}{Logarithmic Model} &
\multicolumn{2}{|c|}{Exponential Model} \\ \hline
Scalar Real & \multicolumn{2}{l|}{$SR = -570 \mp 307\epsilon + 23.1ln(T)$} &
\multicolumn{2}{l|}{$SR = (-491 \mp 362\epsilon)exp(-T/5440)$} \\
& $r_{0} = 1.105$ & $a = 0.692$ & $r_{0} = 1.102$ & $a = 0.700$ \\ \hline
Scalar Imag. & \multicolumn{2}{l|}{$SI = 237 \mp 71.1\epsilon - 42.0ln(T)$} &
\multicolumn{2}{l|}{$SI = (52.6 \mp 125\epsilon)exp(-T/164.2)$} \\
& $r_{0} = 1.157$ & $a = 0.512$ & $r_{0} = 1.153$ & $a = 0.488$ \\ \hline
Vector Real & \multicolumn{2}{l|}{$VR = 532 \pm 235\epsilon -37.4ln(T)$} &
\multicolumn{2}{l|}{$VR = (399 \pm 287\epsilon)exp(-T/1686)$} \\
& $r_{0} = 1.109$ & $a = 0.664$ & $r_{0} = 1.105$ & $a = 0.676$ \\ \hline
Vector Imag. & \multicolumn{2}{l|}{$VI = -189 \pm 54.2\epsilon + 28.9ln(T)$} &
\multicolumn{2}{l|}{$VI = (-54.8 \pm 60.4\epsilon)exp(-T/512.2)$} \\
& $r_{0} = 1.149$ & $a = 0.633$ & $r_{0} = 1.137$ & $a = 0.647$ \\ \hline
\end{tabular}
\end{center}
 
\noindent Figures 3 and 4 show fits to p + $^{208}$Pb data at 200 MeV
using the
proton-only potentials and the neutron-plus-proton potentials (Table I) for
both the logarithmic and exponential models. As can be seen, the fits are
quite good for both models for both input data sets. In fact, on the basis
of this 200 MeV proton data, one cannot determine the preferred model
and there appears to be only a slight preference for the neutron-plus-proton
input data over the proton-only input data. However, the fits to the n +
$^{208}$Pb total cross section data for the identical two models and identical
two input data sets, Fig. 5, show only qualitative agreement in the case of
the proton-only input data whereas quite good agreement is obtained in the
case of the neutron-plus-proton input data. In addition, the data indicate
a slight preference for the logarithmic model over the exponential model.
Furthermore, the predictive power of the identical two models and identical
two input data sets is tested against n + $^{208}$Pb differential elastic
cross section and analyzing power data at 155 MeV \cite{Ha58} that were not
included in the input data. Figure 6 shows that both potentials give similarly
good predictions, but that the analyzing power data clearly prefers the
neutron-plus-proton input data. Thus, Figs. 3--6 lead to the conclusions
that a medium-energy phenomenological {\it nucleon-nucleus} potential may be
the best way to
proceed and (somewhat weaker) that an isoscalar logarithmic energy dependence
and an
isovector energy independence may be more physical than an exponential
energy dependence for both isoscalar and isovector components.
 
\vspace{10pt}
\noindent Given these conclusions the logarithmic model (Table I) was
used to predict unmeasured neutron elastic scattering angular distributions,
analyzing powers, and spin-rotation functions at 100, 200, and 300 MeV
\cite{KM90}.
These are shown in Fig. 7 for 100 MeV as are the corresponding predictions
for proton scattering also using the logarithmic model.
The differences between the three observables for neutron and proton
scattering,
at 200 and 300 MeV as well as 100 MeV, were studied by also performing
calculations
for a ``gedanken'' projectile with potential strengths appropriate to a
proton,
but with the charge set to zero. The study concluded that (a) the shift in
the first
minimum of the differential cross sections is due to the influence of the
Coulomb
interaction, while the enhanced magnitude of the back-angle neutron cross
sections
results from the difference in sign of the isovector strengths, (b) the
saturation
of the neutron analyzing powers (+1.0) appears to come solely from the absence
of
the Coulomb interaction, and (c) the damping of the large-angle oscillations
of
the neutron spin-rotation functions largely arises from the difference in sign
of the isovector strengths, although the absence of the Coulomb interaction
plays some role.
 
\vspace{10pt}
\noindent Finally, the correlations and ambiguities found in Dirac
phenomenology
were studied \cite{KM93} for a single case, that of p + $^{40}$Ca at 181 MeV.
Briefly, two equivalent families of potentials are found, only one of which
predicts the correct total reaction cross section (the measured value was
{\it not}
used in determining the best-fit parameterization), and has a just slightly
lower $\chi^{2}$ than that of the minimum in the other family.
As one might expect, relatively large ambiguities are found in the imaginary
strengths and they are linearly correlated. Also, the real geometries
are particularly stable and the real strengths are also correlated, but are
much better determined than the imaginary strengths. The point to understand
is that the observed total reaction cross section is able to distinguish
the correct Dirac phenomenological potential family.
 
\begin{center}
{\bf A Global Phenomenological Schr\"odinger Potential} \\
\end{center}
\noindent The potential described in this section consists of a global
medium-energy
nucleon-nucleus phenomenological relativistic Schr\"odinger potential. The
potential
is global in projectile energy, projectile isospin, and target (Z,A).
It employs relativistic kinematics and a relativistic equivalent to the
Schr\"odinger equation obtained by appropriate reduction of the Dirac equation
for a massive energetic fermion ($m,\:k$) moving in a localized central
potential $V(r)$ taken as the time-like component of a Lorentz four-vector.
The resultant radial equation for the partial wave $f_{L}(\rho)$
is given by ($\hbar = c = 1$)
\begin{equation}
\left\{\frac{d^{2}}{d\rho^{2}} + \left[1 - \frac{U(\rho)}{T_{c}} -
\frac{L(L+1)}
{\rho^{2}}\right]\right\}\:f_{L}(\rho) = 0
\end{equation}
where $\rho = kr$, $T_{c}$ is the total c.m. kinetic energy, $L$ is the
orbital
angular momentum, and $U(\rho)$ is the
renormalized total (nuclear plus Coulomb) optical potential
\begin{equation}
U(\rho) = \gamma\:V(r)\;,\; \gamma = 1 +\:\frac{T_{c}}{T_{c} + 2m} \; .
\end{equation}
\noindent Equation (19) is formally identical to the radial equation
for the solution of the
non-relativistic Schr\"odinger equation for the analogous scattering problem.
By way of example, Fig. 8 shows calculations of the proton total reaction
cross
section for p + $^{27}$Al using Eq. (19) in three different ways for the
identical potential
$V(r)$ : (1) non-relativistic (classical kinematics and $\gamma \equiv 1$),
(2) relativistic kinematics (and $\gamma \equiv 1$), and (3) relativistic
equivalent Schr\"odinger (relativistic kinematics and $\gamma > 1$). Clearly,
the
$\gamma$ factor becomes increasingly important as the projectile kinetic
energy
increases [Eq. (20)]. Option (3) is used in the remainder of this section.
 
\vspace{10pt}
\noindent The starting point for determining this potential was the
phenomenological proton optical-model potential of Schwandt {\it et al.}
\cite{Sc82} based upon differential elastic scattering cross sections and
analyzing powers for the mass range $24\:\leq\:A\:\leq\:208$ and proton
laboratory kinetic energy range $80\:\leq\:T_{p}\:\leq\:180$ MeV.
The potential employs standard Woods-Saxon form factors. The goals
were to extend the mass range of the potential to $12\:\leq\:A\:\leq\:208$,
to extend the energy range of the potential to $50\:\leq\:T_{p}\:\leq
\:400$ MeV, and to transform the extended proton potential to a neutron
potential for the same mass and energy ranges. Moreover, optimal reproduction
of
the measured integrated scattering observables, the proton total reaction
cross section $\sigma_{R}$ and the neutron total cross section $\sigma_{T}$,
was the main focus of the work.
 
\vspace{10pt}
\noindent The approach used was to (a) adjust {\it only} the parameters of the
proton central absorptive potential to optimally reproduce the measured total
reaction cross sections, (b) perform these adjustments allowing only
{\it small}
changes in the calculated $d\sigma/d\Omega$ and $A_{y}(\theta)$, and (c)
transform the extended proton potential to the corresponding neutron potential
by use of the Lane model \cite{LA62} and accounting for the Coulomb
correction.
[Since the proton starting potential \cite{Sc82} does not explicitly contain a
Coulomb correction term it is assumed that the term is implicitly present and,
therefore, that it must be {\it subtracted} from the corresponding
{\it neutron} potential.
The correction is taken as $0.4\:Z/A^{\frac{1}{3}}$.]
The work was performed by iterative computation, that is, a generalized
nonlinear
least-squares adjustment algorithm was {\it not} used,\protect\footnote{For
this reason
the results have not been submitted for publication in a refereed journal.}
for three nuclei spanning
a large mass range: $^{27}$Al, $^{56}$Fe, and $^{208}$Pb.
The resultant potential gave reasonably satisfactory predictions for both
proton and neutron scattering observables for other target $A$ values in the
same range \cite{MA88}. Further iterative computations were performed for
six additional nuclei: $^{12}$C, $^{16}$O, $^{40}$Ar, $^{81}$Br, $^{107}$Ag,
and $^{138}$Ba. The nine total extracted values of the imaginary diffuseness
parameter $a_{I}$, for Region II of the potential, were then fit by an
expansion
in powers of $A^{\frac{1}{3}}$ as shown in Fig. 9 \cite{AS96}.
With this result, the current parameterization of the potential is given in
Table II.
An example using the potential is given in Fig. 10 for the integrated
observables
of $^{56}$Fe and where ``Modified potential'' refers to Table II.
 
\begin{center}
{\bf Lessons from the Construction and Use of the Two Potentials} \\
\end{center}
\noindent Several conclusions (some of them tentative) come from the work
summarized above. First, the medium-energy phenomenological optical potential
is very forgiving, just like the low-energy phenomenological potential. In
particular, several different projectile energy dependencies appear tractable
(linear, logarithmic, exponential, \ldots) provided the total energy range
is not excessive. Also, in a Schr\"odinger phenomenology, relatively small
adjustments can be made in the parameters
of the absorptive potential to improve agreement with the integrated
observables
without catastrophic consequences for the differential elastic and
spin-dependent
observables. In addition, it appears possible to obtain approximately
smooth \\
 
\begin{center}
\bf{Table II. Schr\"odinger global optical potential for nucleon--nucleus
scattering \\
in the target mass range 12 $\mbox{\boldmath $\leq\:A\:\leq$}$ 208 and in the
projectile \\ energy range 50 $\mbox{\boldmath $\leq\:T\:\leq$}$ 400 MeV.
\protect
\footnote{Strengths are in MeV and geometry is in fm; the upper (lower)
signs refer to neutrons (protons); $\tau_{3}$ is defined
just below Eq. (3).}} \\
\vspace{12pt}
\begin{tabular}{|l|l|} \hline
Real Central & $V_{R} = 105.5[1 - 0.1625ln(T)] \mp 16.5[(N - Z)/A] -
(\frac{1}{2}
+ \tau_{3})(0.4Z/A^{\frac{1}{3}})$ \\ & $r_{R} = 1.125 + T/10^{3}\;,\;
T \leq 130 $ \\ & $ r_{R} = 1.255\;,\; T > 130 $ \\ & $ a_{R} = 0.675 +
3.1T/10^{4}$
\\ \hline
Imag. Central & $W_{V} = 6.6 + 2.73(T - 80)/10^{2} + 3.87(T - 80)^{3}/10^{6}
\;,
\; T \leq 140$ \\ & $W_{V} = 7.314 + 0.0462T \;,\; T > 140$ \\ & $r_{I} = 1.65
- 2.4T/10^{3} \;,\; T \leq 140$ \\ & $r_{I} = 1.17 \;,\; T > 140$ \\ &
$a_{I} = 0.27 + 2.5T/10^{3} \;,\; T \leq 140$ \\ & $a_{I} = 0.3537 +
0.08451A^{\frac{1}{3}} - 0.001835A^{\frac{2}{3}} \;,\; T > 140$ \\ \hline
Real Spin-Orbit & $V_{SO} = 19.0[1 - 0.166ln(T)] \pm 3.75[(N - Z)/A]$ \\ &
$r_{VSO} = 0.920 + 0.0305A^{\frac{1}{3}}$ \\ & $a_{VSO} = 0.768 - 0.0012T \;,
\; T \leq 140$ \\ & $a_{VSO} = 0.60 \;,\; T > 140$ \\ \hline
Imag. Spin-Orbit & $W_{SO} = 7.5[1 - 0.248ln(T)]$ \\ & $r_{WSO} = 0.877
+ 0.0360A^{\frac{1}{3}}$ \\ & $a_{WSO} = 0.62$ \\ \hline
\end{tabular}
\end{center}
energy dependencies of the various
observables with piecewise (continuous and discontinuous)
parameterizations. [Note that this is not always the case
for the transmission coefficients.]
The Schr\"odinger phenomenology of Table II contains such parameterizations
because the starting potential \cite{Sc82} was constructed in this way.
In general, however, piecewise parameterization should clearly be avoided.
Second, a Dirac phenomenology may provide physically realistic potentials
over a wider projectile energy range than a Schr\"odinger phenomenology
because the effective central potential in a second-order reduction involves
squares and cross terms of the form factors appearing. This allows, for
example, a ``wine-bottle'' shape. Third, in a Schr\"odinger phenomenology
employing Woods-Saxon form factors, it appears that small adjustments in
the imaginary diffuseness parameter $a_{I}$ can fine tune the integrated
observables with a minimal impact on the other observables. Perhaps the same
is true in Dirac phenomenology? Fourth (and last), given the extreme
sparseness of experimental medium-energy neutron scattering differential
elastic and
spin-dependent observables, a Dirac or relativistic Schr\"odinger
phenomenological approach that is global in
(1) projectile energy, (2) projectile isospin, and (3) target (Z,A),
and uses the existing medium-energy proton and neutron (total cross sections)
databases, appears
to be a tractable way to calculate physically realistic neutron elastic
scattering observables over wide ranges in energy and target.
Whether a Dirac or relativistic Schr\"odinger formalism should be used has
yet to be determined.
\newpage
 
\begin{center}
{\bf Current Work} \\
\end{center}
\noindent Because a satisfactory global medium-energy nucleon-nucleus optical
potential does not yet exist we are continuing our work on this goal.
Currently, we are addressing an energy range of (perhaps) 20 MeV to
(perhaps) 2000 MeV and a (spherical) target mass range of 16 to 209.
The experimental database (currently over 20000 points) consists of
the sets \{$d\sigma/d\Omega, \sigma_{R}, A_{y}, Q$\} for protons and
\{$\sigma_{T}$, some $d\sigma/d\Omega$ and $A_{y}$\} for neutrons. Our
approach
is to consider both relativistic Schr\"odinger and Dirac phenomenology with
the {\it identical} database in a nonlinear least-squares adjustment
algorithm. Piecewise parameterizations will be inadmissable.
We will also address a microscopic Dirac approach for the same ranges
employing
proton and neutron densities from recent work on the nuclear bound state
problem using a relativistic Hartree approach \cite{NHM92}. Here, only
even-even target nuclei will be considered.
This topic is particularly exciting because the relativistic Hartree approach
that we use is easily extended to relativistic Hartree-Fock \cite{NHM92} and,
perhaps more importantly, we have discovered that our coupling constants
are mostly {\it natural} (of order unity) when our Lagrangian is rewritten
in a form that is based upon QCD scaling and chiral symmetry \cite{FML96}
and whose validity demands {\it naturalness}.

\newpage
\begin{center}
{\bf Figure Captions} \\
\end{center}
 
\begin{description}
\item[Fig. 1] $\chi^{2}$/point/data set for p + $^{208}$Pb scattering in the
energy range 80--800 MeV for four energy dependencies.
\item[Fig. 2] $\chi^{2}_{tot}$/point for p + $^{208}$Pb scattering in three
energy ranges for three energy dependencies.
\item[Fig. 3] Differential elastic cross sections for p + $^{208}$Pb
scattering
at 200 MeV in the logarithmic and exponential models. The solid curves are
obtained in calculations that use the Dirac global potential (Table I)
which has been determined by simultaneously fitting neutron and proton data.
The dashed curves are from calculations using a Dirac global potential in
which only proton data have been fit.
\item[Fig. 4] Spin observables for p + $^{208}$Pb scattering at 200 MeV in
the logarithmic and exponential models. The solid and dashed curves have
the same explanation as in Fig. 3.
\item[Fig. 5] Total cross sections for n + $^{208}$Pb scattering from 95 to
250 MeV in the logarithmic and exponential models.
The solid and dashed curves have the same explanation as in Fig. 3
and the dotted curve is the prediction of a geometric black disk model.
\item[Fig. 6] Differential elastic cross sections and analyzing powers for
n + $^{208}$Pb scattering at 155 MeV in the logarithmic and exponential
models. The solid and dashed curves have the same explanation as in Fig. 3.
The calculations shown are {\it predictions} as the experimental data were
not used in determining the potentials.
\item[Fig. 7] Differential elastic cross sections and spin observables for
nucleon-plus-$^{208}$Pb scattering at 100 MeV as predicted by the Dirac
global logarithmic potential (Table I). The solid curves are the predictions
for neutron scattering while the dashed curves are the predictions for
proton scattering.
\item[Fig. 8] Calculations of the proton total reaction cross section
for the p + $^{27}$Al reaction using the Schr\"odinger formalism, Eq. (19),
in three different approaches with the identical optical potential.
\item[Fig. 9] Empirical values of and polynomial fit to the imaginary
diffuseness
parameter $a_{I}$ for Region II ($T_{p}\:>\:140$ MeV) of the Schr\"odinger
global potential of Table II.
\item[Fig. 10] Comparisons of measured and calculated integral scattering
observables
from the nucleon-plus-$^{56}$Fe reaction using the Schr\"odinger formalism.
The original potential is that of
Schwandt {\it et al.} \cite{Sc82} and the modified potential is that of
Table II.
\end{description}
\end{document}